# Charge-Density-Wave Proximity Effects in graphene

Boram Kim[1†], Jeehoon Park[1†], Jinshu Li[2†], Hongsik Lim[1], Gyuho Myeong[1], Wongil Shin[1], Seungho Kim[1], Taehyeok Jin[1], Qi Zhang[2], Kyunghwan Sung[1], Kenji Watanabe[3], Takashi Taniguchi[3], Euyheon Hwang[2**], Sungjae Cho[1*]

**The proximity effects of superconductivity, ferromagnetism, and spin-orbit coupling in semiconductors and metals have been extensively studied. However, the properties of charge density wave (CDW) systems combined with other electronic materials have not yet been investigated. Here, we incorporate the CDW properties of 1T-TaS$_2$ into the electronic transport in graphene for the first time. During the CDW phase transitions, anomalous transport behaviors were observed within the graphene, which are closely related to the formation of correlated disorder in 1T-TaS$_2$. In particular, the commensurate CDW forms a periodic charge distribution with potential fluctuations, and thereby constitutes the correlated charged impurities, decreasing resistivity and enhancing carrier mobility in graphene. The demonstration of the CDW-graphene heterostructure system paves a way to control the temperature-dependent carrier mobility and resistivity of graphene and to develop novel functional electronic devices such as graphene-based sensors and memory devices.**

The layered transition metal dichalcogenide (TMDC) 1T-TaS$_2$ shows successive charge density wave (CDW) phases at different temperatures[1–3]. At high temperature of over 550 K, the ground state is metallic. Upon cooling, 1T-TaS$_2$ undergoes a series of first-order phase transitions to an incommensurate CDW (ICCDW) phase (550K) to a nearly commensurate CDW (NCCDW) phase (350K), and finally to a commensurate CDW (CCDW) phase at the lowest temperature (180K). As the temperature decreases, 12 outer Ta atoms gather to the 13$^{th}$ center Ta atom and form the shape of star-of-David. CDW phases of 1T-TaS2 are distinguished by commensuratability of this star-of-David cluster to the underlying lattice. In ICCDW phase, Ta atoms deviate little from their original location but don't form the star-of-David clusters. Under 350K, star-of-David starts to generate. They make domains consisting of star-of-David with discommensurate CDW boundaries. Under 180K, all space are filled with star-of-David. Especially in the CCDW phase, the star-of-David cluster consisting of 13 Ta atoms forms a regular lattice structure[1,4–7]. The electron in the cluster's central atom is localized, leading to an insulating Mott state[1,2,8–10]. Upon cooling from a high temperature ICCDW phase to the NCCDW and CCDW phases, a series of transitions occur exhibiting hysteretic transport behavior. Significant experimental efforts have been made to understand the nature of these rich CDW phases of 1T-TaS$_2$ including the effect of doping[11,12], pressure[13] or substrate[14], STM and TEM imaging of each CDW phase. The low temperature CCDW phase disappears and new "hidden states" appear upon perturbations such as electrical currents and pulse[15] or laser pulse irradiation[16,17]. The applications within memory devices and sensors (such as light and heat) have been explored by employing the hysteresis property and collective excitation [17–20]. Although the atomic structures and electronic conductions of each CDW phase have been widely studied, the effects of the CDW phases to other adjacent electronic systems in a combined heterostructure have not yet been studied.

[1] Department of Physics, Korea Advanced Institute of Science and Technology (KAIST), Daejeon, Korea

[2] SKKU Advanced Institute of Nano Technology and Department of Nano Engineering, Sungkyunkwan University, Suwon, Korea.

[3] National Institute for Materials Science, Namiki Tsukuba Ibaraki 305-0044, Japan

[†]These authors contributed equally to this work
* Corresponding author, S. C, Email: sungjae.cho@kaist.ac.kr

** Corresponding author, E. H., Email: euyheon@skku.edu

As graphene trigger enormous interest since 2005, mechanical exfoliation, transport theory of graphene characterized by Dirac fermion have studied successfully. Through active research, E H Hwang et al. compile the theory of transport of graphene in case with the diverse scattering source types[21]. Among many scatterers, they found that the resistivity of graphene on charged impurity become low when the charged impurities are correlated[22,23].

Here, we report CDW proximity effects in graphene for the first time. We unexpectedly observe anomalous transport behaviors in graphene depending on the CDW phases of TMDCs. The lattice distortion and charge redistribution by the spatial distribution of star-of-David units in 1T-TaS$_2$ act as a source of the charged impurities in graphene. Strikingly, we find that when 1T-TaS$_2$ undergoes a transition from the metallic NCCDW phase to the insulating CCDW phase, the resistivity of graphene decreases abruptly at the transition temperature. Simultaneously, the field-effect mobility of graphene also shows a sudden change at the transition temperature. This abrupt change in electronic transport of graphene at the CDW transition temperature can be attributed to the charge redistribution within 1T-TaS$_2$. The stars-of-David are commensurately distributed in the CCDW phase. Hence, correlations arising in the spatial distribution of the disorder lead to the sudden decrease in graphene resistivity upon the transition from NCCDW phase to CCDW phase in 1T-TaS$_2$. Conversely, because the NCCDW phase consists of insulating star-of-David domains with discommensurate CDW boundaries, the correlation decreases and the domains of the star-of-David cluster provide an additional scattering source to the graphene. Therefore, the graphene resistivity increases suddenly upon the transition from the ICCDW phase to the NCCDW phase within 1T-TaS2. Finally, in ICCDW phase at high temperatures, charged impurities arising from randomly distributed stars-of-David disappear, which gives rise to the lower resistivity in ICCDW than that in NCCDW. This anomalous temperature-dependent electronic transport in graphene results from proximity to charge density wave phases in 1T-TaS$_2$ and the sudden change of electronic transport in graphene could be applied to the development of new functional electronic devices, such as graphene-based sensors and memory devices.

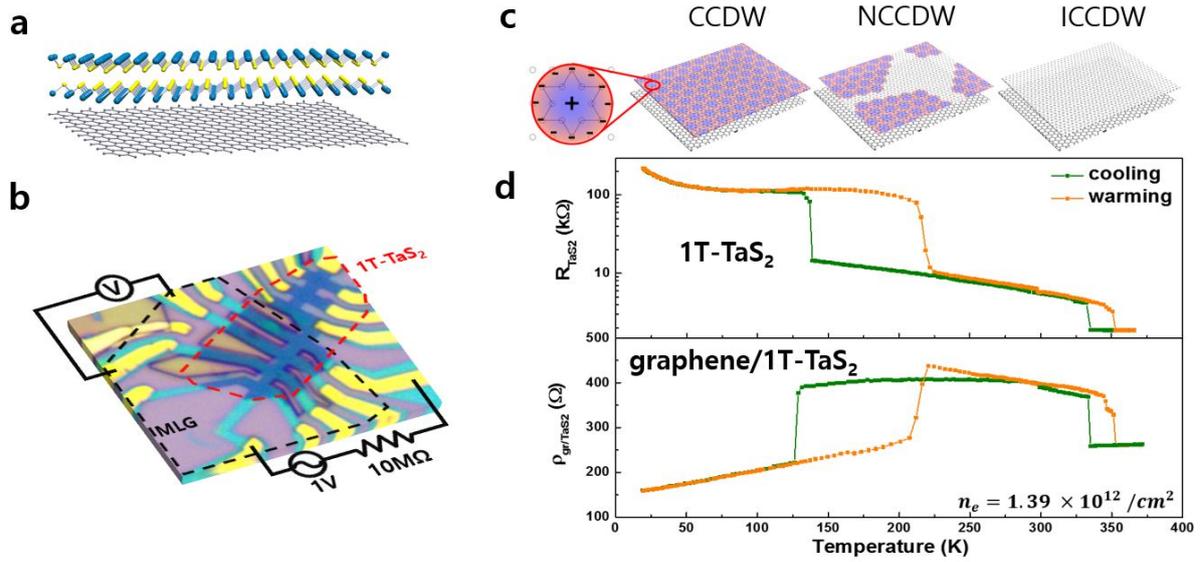

**Fig. 1. Schematic of the heterostructure, an optical image, and characteristic temperature-dependent transport curves of the device. a.** Schematic image of the monolayer graphene under 1T-TaS$_2$. **b.** Optical image of the graphene/TaS$_2$ heterostructure etched in the Hall bar pattern. Red and black lines denote the 1T-TaS$_2$ and monolayer graphene, respectively. Four-probe measurement connection is illustrated. **c.** Schematic image of the 1T-TaS$_2$ structure over the graphene layer with different temperature-dependent phases; CCDW (left)–NCCDW (middle)–ICCDW (right). **d.** Temperature dependence of the resistance for 1T-TaS$_2$ (upper graph) and resistivity of the graphene/1T-TaS$_2$ heterostructure (lower graph). Green (orange) line represents the process of cooling (warming).

The device channels consist of monolayer graphene and few layer 1T-TaS$_2$ (Fig. 1a). A thin layer of 1T-TaS$_2$ does not undergo the NCCDW-CCDW phase transition due to thinning-induced slow ordering kinetics[24,25]. Thus, we used a 1T-TaS$_2$ layer with a thickness of 3–8 nm. Van der Waals epitaxy was performed inside an N$_2$-filled glovebox until the heterostructure was encapsulated by hexagonal boron nitride (hBN) to avoid contamination from the air exposure or chemicals especially in the interface of graphene and 1T-TaS$_2$ (see Supplementary Fig. S2). The graphene/1T-TaS$_2$ heterostructure was etched in the Hall bar pattern and contacted by Cr/Au electrodes for electrical measurement. The device consists of two regions, one with only 1T-TaS$_2$ in contact with the metal, and the other with graphene lying beneath the 1T-TaS$_2$, as shown in Fig. 1b. Each part of the devices was characterized separately by performing measurements at different temperatures and gate voltages. To apply gate voltages, we used hBN layers of a thickness of 10–20 nm and 285 nm SiO$_2$ as the gate dielectrics (see Supplementary Fig. S3).

Figure 1d shows resistance as a function of temperature for the two (1T-TaS$_2$ only and 1T-TaS$_2$/graphene) regions of the devices across the ICCDW-NCCDW-CCDW phases of 1T-TaS$_2$, as shown in the schematics of Fig. 1c. The CDW transitions depend on the cooling/warming rate, which was set to 0.2 K/min or lower for the measurement of all the devices in this study. The temperature-dependent resistance of the 1T-TaS$_2$ in the top panel of Fig. 1d shows abrupt increases upon transitions from the ICCDW to NCCDW phase and from the NCCDW to CCDW phase. The hysteresis loop between cooling and warming indicates the metastability between each phase (1$^{st}$ order transition). Surprisingly, the graphene/1T-TaS$_2$ resistivity shows abrupt changes at each CDW transition temperature of 1T-TaS$_2$ with the same hysteresis loops, as shown in the bottom panel of Fig. 1d. The resistivity of graphene/1T-TaS$_2$ increases during the ICCDW–NCCDW transition of 1T-TaS$_2$, while it decreases during the NCCDW–CCDW transition with a corresponding increase in the resistance of 1T-TaS$_2$. This indicates that the changes in the resistance of graphene/1T-TaS$_2$ region do not originate from 1T-TaS$_2$, but from graphene.

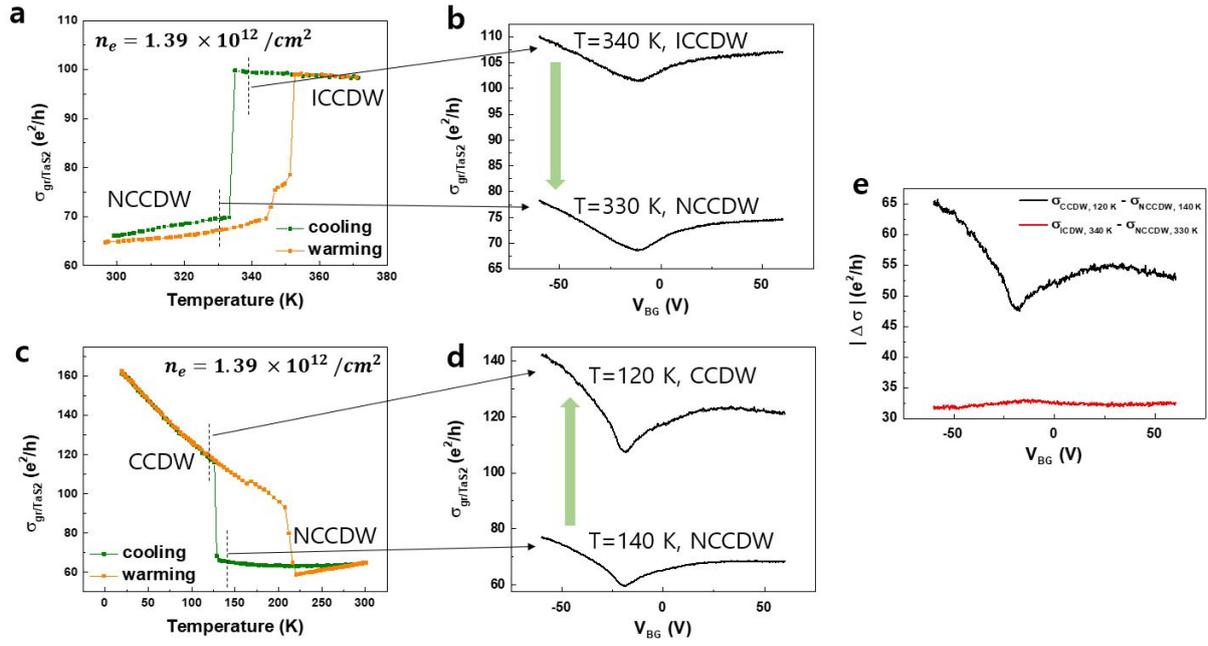

**Fig. 2. Gate- and temperature-dependent conductivity curves of the 1T-TaS$_2$/graphene heterostructure. a**. Conductivity within the range of room temperature to high temperature for the cooling (green) and warming (orange) process at V$_g$ = 0 V. (carrier density $n_e = 1.39 \times 10^{12}/cm^2$) **b.** Gate-dependent conductivity curves before (340 K) and after (330 K) the phase transition from ICCDW to NCCDW, during the cooling process. **c.** Conductivity within the range of low temperature to room temperature for cooling (green) and warming (orange) process in the V$_g$ = 0 V. (carrier density $n_e = 1.39 \times 10^{12}/cm^2$) **d.** Gate-dependent conductivity curves before (140 K) and after (120 K) the phase transition from NCCDW to CCDW, during the cooling process. **e.** Gate-dependent conductivity difference passing through the phase transition, from NCCDW to CCDW (black line) and from ICCDW to NCCDW (red line).

Figure 2 shows the gate-dependent conductivity of graphene/1T-TaS$_2$ across the CDW transitions in 1T-TaS$_2$. During the ICCDW-NCCDW transition (Fig 2a, b), the conductivity of the heterostructure decreases abruptly. The gate-dependent conductivity shows the Dirac point of graphene approximately $V_g = -19.8$ V. Note that 1T-TaS$_2$ shows negligible gate-dependent resistance at all temperatures (see Supplementary Fig. S5). With graphene dominating the gate-dependent conductivity of the heterostructure, sudden decreases of conductivity occur at all gate voltages, indicating that the ICCDW–NCCDW transition affects electronic transport in graphene regardless of the carrier type. As shown in Fig. 2e, the magnitude of the decrease in conductivity varies across the ICCDW–NCCDW transition and shows negligible gate-dependence. During the NCCDW–CCDW phase transition (Fig 2c, d), the conductivity of the heterostructure increases abruptly with the abrupt increase in the resistivity of the 1T-TaS$_2$ flakes in the heterostructure, due to the Mott insulating state of the CCDW phase (Fig. 1d). This indicates that the change in conductivity of the graphene/TaS$_2$ heterostructure during the NCCDW–CCDW transition is not dominated by the conductivity of 1T-TaS$_2$, but by that of graphene. Even though 1T-TaS2 has comparable resistance to the graphene in NCCDW phase, if we consider 1T-TaS2/graphene region as a resistivity in parallel problem, we can find the resistivity of graphene decreases at the transition from the NCCDW to CCDW phase, in the opposite way of 1T-TaS2 (see Supplementary Fig. S7). The gate-dependent conductivity of the heterostructure shows that this sudden increase in conductivity upon transitioning into the CCDW phase occurs at all gate voltages, regardless of electron or hole conduction. The magnitude of the increase in conductivity across the NCCDW–CCDW transition does not change significantly with the gate voltage, as depicted by the black curve in Fig. 2e.

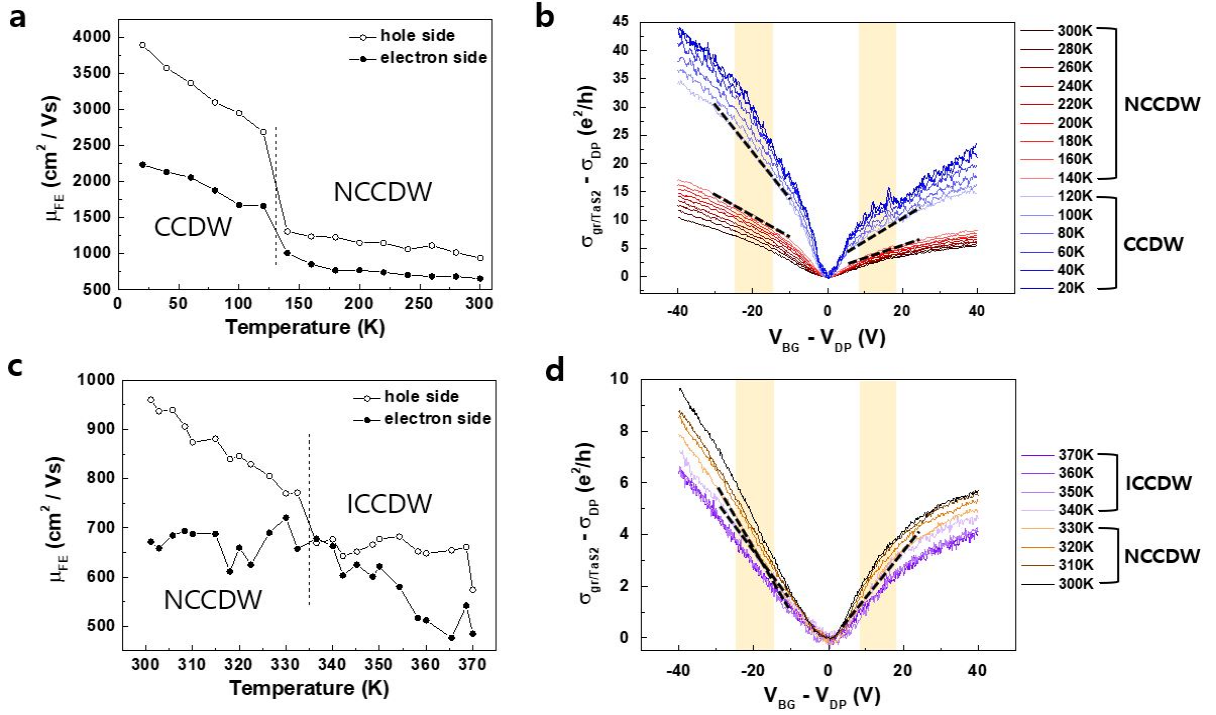

**Fig. 3. Temperature dependent field-effect mobility of 1T-TaS$_2$/graphene. a, b.** Field effect mobility change with temperature range from 300 K to 20 K (a) and from 370 K to 300 K (c) during the cooling process. **b, d.** Gate dependent conductivity subtracted by the conductivity value at the Dirac point for every curve. For the NCCDW-CCDW transition (b), a sudden increase in mobility occurs, unlike in the ICCDW-NCCDW transition (d). Yellow gate voltage regions are used to calculate the field effect mobility and black dashed lines guide the slope of gate-dependent conductivity which is proportional to mobility.

Figure 3 shows the temperature-dependent field-effect mobility of the heterostructure. Upon the transition from the NCCDW to CCDW phase of 1T-TaS$_2$ during cooling at approximately T = 130 K, the field-effect mobility increases abruptly from 1300 cm$^2$/Vs to 2700 cm$^2$/Vs. Conversely, a special mobility change in the heterostructure is not observed during the ICCDW–NCCDW transition.

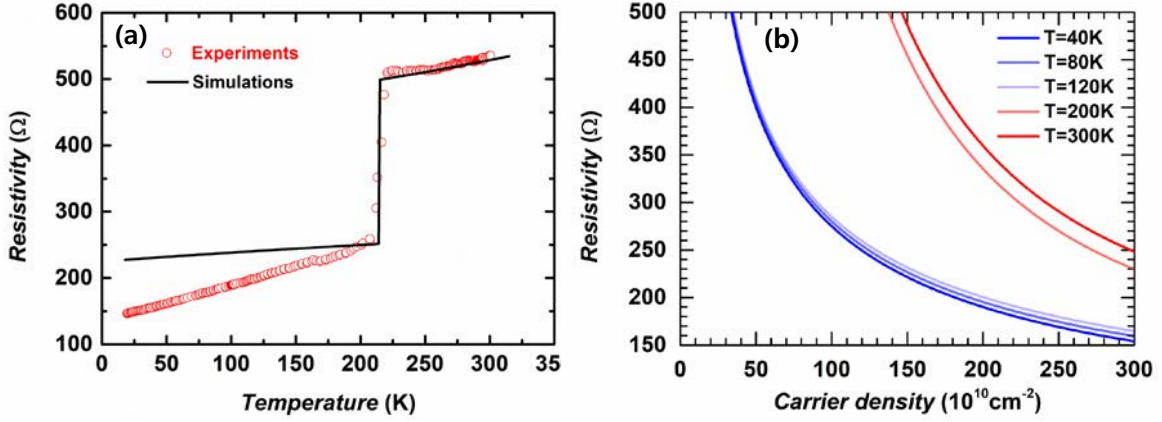

**Fig. 4. Calculated graphene transport of 1T-TaS$_2$/graphene system. a.** Graphene resistivity at a carrier density of $n = 1.39 \times 10^{12}$ cm$^{-2}$ as a function of temperature. **b.** Calculated graphene resistivity as a function of carrier density for various temperatures. The red (blue) curves represent the conductivities in the NCCDW (CCDW) phase of 1T-TaS$_2$. Due to the correlation effect of the star-of David, the graphene resistivity shows a sudden change at the transition temperature between CCDW phase and NCCDW phase.

To verify the origins of the sudden changes in graphene conductivity and mobility during the CDW transitions of 1T-TaS$_2$, we perform theoretical calculation. Figure 4 presents calculated results based on the Boltzmann transport theory[21]. In the calculation we fully consider the well known scattering sources in graphene transport (i.e., charged disorder with temperature-dependent screening and phonon scattering effects are considered). We provide theoretical details in the supporting materials. The theory reflects the calculated transport results of graphene after taking into account the additional carrier scattering arising from the CDW phase transition of 1T-TaS$_2$. According to the CDW phases, the lattice distortion of 1T-TaS$_2$ produces charged impurities, whose configuration affects the transport behavior of Dirac fermions in graphene. Figure 4(a) shows the graphene resistivity as a function of temperature at a density of $n = 1.39 \times 10^{12}$ cm$^{-2}$, which fits to the theoretical calculation.

At high temperatures ($T > 330$ K), 1T-TaS$_2$ is in the ICCDW phase and no structural distortion into star-of-David is expected. In the ICCDW phase, 1T-TaS$_2$ consists of a triangular lattice of Ta atoms sandwiched between two triangular lattices of S atoms. The resulting triangular lattice is charge-neutral. Thus, in the combined system of graphene and TMDC, TMDC plays a role in reducing overall mobility of graphene because of the interface disorders and the polar optical phonons at high temperatures, which is considered in the calculation.

Upon transition from the ICCDW to NCCDW phase, the domains of the star-of-David are formed. The star-of-David distortion consists of 13 Ta atoms, where 12 Ta ions are attracted to each other and produce a CDW bandgap in the CCDW phase. The electron at the central atom in the cluster is localized, leading to an insulating Mott state. The Ta ions in the stars behave as charged impurities and provide new scattering sources in the graphene carrier transport. Although charged impurities are formed during the transition from the ICCDW phase to NCCDW phase, the carrier transport in graphene at this high temperature regime is dominated by phonon scattering. Therefore, the mobility change in graphene is negligible during the ICCDW–NCCDW transition within 1T-TaS2. However, the domains of the star-of-David produce random charged impurities in the NCCDW phase. The resistivity of graphene is expected to increase when random charged impurities are introduced to graphene. As shown in Fig. 4(a), a sudden increase in resistivity at the transition point (T=330 K) is observed.

As the temperature decreases further, we observe another CDW phase transition, i.e., the NCCDW-CCDW phase transition, of 1T-TaS$_2$ at a low temperature (T=130 K). At this transition, the clusters of the stars-of-David in the NCCDW phase merge to form a system size of lattice of the stars, i.e., periodically arranged star-of-David. Thus, the charge impurities of stars have some spatial correlations in the distribution of the charged impurity locations within the system, i.e., these charged impurities are no longer considered to be completely random in the spatial domain. The impurity correlation effects on the graphene transport have been considered in Refs. [22,23] and it is found that the spatial correlation of impurities gives rise to the reduction of the resistivity. Although local Ta ions are positive, 1T-TaS$_2$ is globally charge-neutral due to the clouding electrons, leading to no additional doping in graphene. The graphene resistivity and field-effect mobility exhibited a sudden decrease and increase, respectively, owing to the correlation effects of the charged stars in 1T-TaS$_2$. The observed sudden drop in graphene resistivity at the NCCDW–CCDW transition temperature can be attributed to the correlation effects of the charged impurities formed during the CCDW phase of 1T-TaS$_2$, as the theoretically calculated results fit the data appropriately.

Figure 4(b) shows the calculated results of graphene/1T-TaS$_2$ resistivity as a function of carrier density at various temperatures. In the calculation we consider all disorder effects discussed before. The red (blue) curves represent the conductivities in the NCCDW (CCDW) phase of 1T-TaS$_2$. Due to the correlation effect of impurities, the graphene conductivities in the CCDW phase show a more sublinear behavior than those in the NCCDW phase.

**Conclusion**

In conclusion, we demonstrated the charge density wave proximity effect on electronic transport in graphene. The van der Waals epitaxy technique allows us to combine these layered TMDCs with graphene. The commensurate CDW forms a periodic charge distribution with potential fluctuations, resulting in correlated charged impurities that affect the carrier transport in graphene. Upon correlation of the charged impurities at the NCCDW–CCDW phase transition in 1T-TaS$_2$, the graphene mobility increases by a few factors. This demonstration of the CDW-graphene heterostructure system introduces a method to control temperature-dependent carrier mobilities and resistivities of graphene; this method could be applied to new functional electronic devices.

# Methods

**Device Fabrication and Measurement**

We first prepared 1T-TaS$_2$, graphene and hBN flakes on a 90 nm SiO$_2$/Si wafer via mechanical exfoliation from bulk crystals in an N$_2$-filled glove box (< 0.1 ppm of H$_2$O and O$_2$) to maintain clean surfaces and prevent contamination from air exposure. Few layer 1T-TaS$_2$ and graphene were identified using the observed optical contrast, Raman spectroscopy, and atomic force microscopy (AFM). Each piece was picked up using a standard dry transfer method[26–28] with a polydimethylsiloxane (PDMS) stamp covered with a polycarbonate (PC) film, and was transferred onto a 285 nm wafer. The PC film was washed with chloroform, acetone, and isopropyl alcohol (IPA). Then, standard e-beam lithography and CF$_4$/O$_2$ plasma etching, followed by e-beam deposition, were conducted to place electrical contacts on the vdW layers. Additional e-beam lithography and metal deposition were performed to place the gate electrode (see Supplementary Fig. S2).

**Measurement**

To obtain the transfer curves, we performed electrical measurements separately from a high temperature to room temperature, and from room temperature to a low temperature (base temperature at T = 20 K) in two home-built measurement vacuum chambers. For temperature-dependent measurements, we maintained a cooling/warming rate of 0.2 K/min or lower. All electrical measurements were performed using standard AC lock-in techniques. (Fig. 1b).

**Theoretical Calculation**

To understand the temperature and density dependence of our devices, we consider the quantitative theory for carrier transport in the presence of all possible disorders arising in the graphene and 1T-TaS$_2$ combined system (i.e., charged impurities, neutral short range disorders, phonons, and additional disorder arising from the phase transition of 1T-TaS$_2$). It is crucial to include the role of disorder from 1T-TaS$_2$ to explain the measured anomalous behavior of graphene transport. Then the total scattering time is calculated by

$$\frac{1}{\tau_t} = \frac{1}{\tau_i} + \frac{1}{\tau_{ph}} + \frac{1}{\tau_{TaS_2}}$$

where $\tau_i$, $\tau_{ph}$, and $\tau_{TaS2}$ are the transport scattering times of impurities (short and long range), all phonons, and disorder from the phase of TaS$_2$, respectively.

**Data availability**

**Acknowledgments**

S. C. acknowledges support from Korea NRF (Grant Nos. 2020M3F3A2A01081899, and 2020R1A2C2100258) ) and E. H. acknowledges support from Korea NRF (Grant No. 2021R1A2C1012176).

**Author contributions**

S. C. conceived and supervised the project. B.K. and J.P. fabricated devices and performed measurements. K.W. and T.T. grew high-quality hBN single crystals. K.M., W.S., S.K., K. S., H.L., and T.J. assisted transport measurements. E.H., J. L., and Q. Z. developed the theoretical model. S. C., B.K., J.P. and E.H. analyzed the data and wrote the manuscript. All the authors contribute to editing the manuscript.

**Competing interests**

The authors declare no competing financial interests.

**Additional information**